\documentclass[times, twoside]{zHenriquesLab-StyleBioRxiv}
\usepackage{blindtext}
\usepackage[colorlinks=true,allcolors=blue]{hyperref}
\leadauthor{Lewis-Douglas} 

\begin{document}

\title{Galvanalyser: A Battery Test Database}
\shorttitle{Galvanalyser}

\author[1,3]{Adam Lewis-Douglas}
\author[2,3]{Luke Pitt}
\author[1,3]{David Howey}

\affil[1]{Battery Intelligence Lab}
\affil[2]{Oxford Robotics Institute}
\affil[3]{Department of Engineering Science, University of Oxford, Parks Road, Oxford, OX1 3PJ, United Kingdom}

\maketitle

\begin{abstract}
Performance and lifetime testing of batteries requires considerable effort and expensive specialist equipment. A wide range of potentiostats and battery testers are available on the market, but there is no standardisation of data exchange and data storage between them. To address this, we present \mbox{\textit{Galvanalyser}}, a battery test database developed to manage the growing challenges of collating, managing and accessing data produced by multiple different battery testers. Collation is managed by a client-side application, the `Harvester', which pushes new data up to a PostgreSQL database on a server. Data access is possible in two ways: firstly, a web application allows data to be searched and viewed in a browser, with the option to plot data; secondly, a Python application programming interface (API) can connect directly to the database and pull requested data sets into Python. We hope to make Galvanalyser openly available soon. If you wish to test the system, please contact us for early access.
\end {abstract}


\begin{corrauthor}
david.howey\at eng.ox.ac.uk
\end{corrauthor}

\section*{Motivation}

Battery testers and potentiostats---manufactured by companies such as MACCOR, BioLogic, BaSyTec, Arbin, Digatron, Ivium and others---are important tools for researchers investigating battery performance and lifetime. Different test systems offer different capabilities, for example some include impedance measurement, and each has bespoke control and data management software. In many labs more than one manufacturer's equipment is in use at the same time. Typically, a battery tester is managed by a local computer which also acts as a data storage device. Therefore adding more items of test equipment not only increases the chances of equipment failure but also means that data is stored in an increasingly fragmented way, spread across several local machines in different bespoke file formats.

Battery cycling experiments can last for weeks, months or even years. During this time both the hardware and software of the tester and of the connected local computer is required to work without failure. But hard drives can malfunction and operating systems or the software running on them can crash, causing possible data loss or corruption. All of this leads to a locally stored dataset becoming increasingly vulnerable as time progresses during long experiments. 

Data access to a fragmented mixture of systems can be challenging, and physical access to battery testers may not always be possible. Machines may be connected to a local network, allowing a remote desktop connection. This however, comes with drawbacks e.g.\ security concerns, and giving researchers unrestricted access to data introduces the possibility of human error, such as accidental deletion. These problems are magnified when multiple systems are in use. Although some systems include a database option for backup, this is not universal.

\section*{Existing approaches}

A variety of software packages have been developed to address these issues over the last decade. For example, commercial solutions such as Voltaiq \cite{noauthor_voltaiq_2020} and Astrolabe Analytics \cite{noauthor_battery_2020} are available. Recently, open source tools such as BEEP \cite{herring_beep_2020} and Samuel Buteau's Universal Battery Database \cite{samuel-buteau_samuel-buteauuniversal-battery-database_2020} have also been released. These open source tools offer various features, mainly focused on data analysis.

\section*{Solution}

\begin{figure*}[t]
\centering
\includegraphics[width=.8\linewidth]{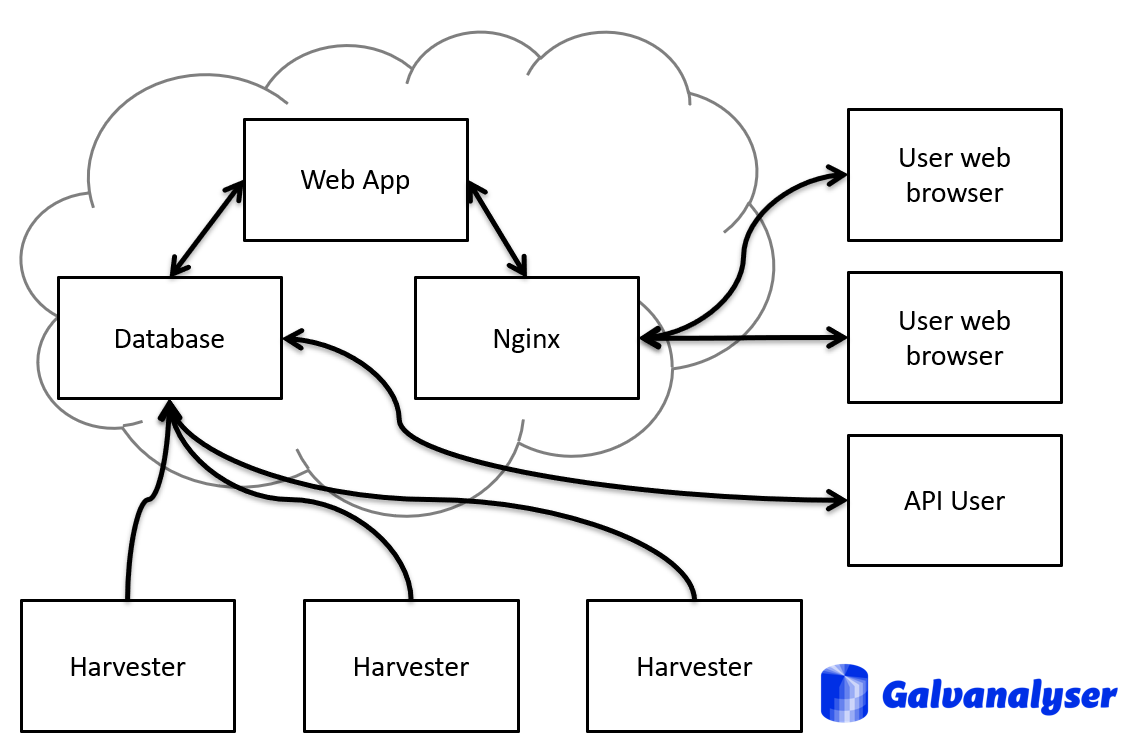}
\caption{Galvanalyser system overview showing client-side `Harvesters', server-side components, and webapp/API}
\label{fig:GalvanalyserSystem}
\end{figure*}

Galvanalyser is a database system for collecting and collating multiple different types of battery test data. Complementary to the approaches described in the previous section, Galvanalyser at present focuses mainly on the initial data acquisition, parsing, and storage step required to deal with a variety of different testers, rather than on downstream data analysis. 

The system consists of two components: a server, and a client, as shown in Fig.\ \ref{fig:GalvanalyserSystem}. The `Harvester' is the client-side application, typically placed on the PC connected to the battery cycler, and this scans selected file-system locations for changes, then uploads these changes to the  server. The Galvanalyser server consists of three parts: (i) a PostgreSQL \cite{group_postgresql_2020} database, which stores and manages all of the battery cycler data; (ii) a web application, which enables querying and plotting of data through a web browser; (iii) an Nginx server, which forms the front-end to the webapp. This approach facilitates data access, providing a simple search interface, as well as having built in data plotting features.

The Galvanalyser system provides a clean solution to the issue of fragmented data storage across many battery cycler systems. With a Harvester application on each local machine, bespoke to that manufacturer, data is backed up as it is produced. The PostgreSQL database creates a centralised and easy to query option for which standard data protection measures can be easily used, such as a RAID system to provide local hardware redundancy, or an offsite backup which provides an added layer of protection.

Long term battery cycling experiments are also given an added layer of protection. Hardware and software faults on attached local machines cannot directly impact previously recorded data in the database, so any issues can be dealt with quickly and without having to worry about potential data loss, giving some peace of mind to users. 

Alongside this, data access is also greatly improved because all equipment-specific data is first parsed and then stored in a common format. The webapp, accessible through a browser, manages access to the PostgreSQL database, and this allows for data from multiple experiments to be compared. In parallel to the webapp, the database is also accessible via a Python API, allowing more complex searching, filtering and analysis.  

In some scenarios data access needs to be restricted and only available to certain users. For example data might be sensitive or protected under a confidentiality agreement. The Galvanalyser database achieves this by requiring user credentials for access, and having dataset access tied to these user credentials. This allows multiple users to share a single database but each only have access to a subset of data, as specified by a system administrator. User access can also be set to read-only if required, which is useful for example for wider sharing of data.

\section*{Galvanalyser technical details}

The Galvanalyser server consists of three elements, each of which is containerised using Docker \cite{noauthor_empowering_2020}. This has several benefits. Containers allow for a program to be combined with all relevant libraries and dependencies and deployed as one package, ensuring that the program will run on any machine, regardless of target machine settings.  

A Docker compose file was created to simplify starting the complete server-side system, including the database, the webapp, and the Nginx server. The client-side element of the application, the Harvester, can be run natively or in a Docker container. This approach is again advantageous because it ensures cross platform compatibility, guaranteeing that the Harvester can run on whatever local machine is connected to a given battery cycler.

\subsection*{The PostgreSQL database}

stores all of the battery cycler data. The database relational structure is explained in more detail in Supplementary Note \ref{appendix}. The data for all types of cycler is stored in a common format in the database. The database also constitutes a single system which can then be backed up easily in many different ways. Visibility of and access to data is restricted using row level security settings.

\subsection*{The web application}

provides access to the database via a browser. It is a Flask-based Python3 application that uses the Dash framework, which is based on React. It uses Plotly for plotting data and WebGL acceleration where available. The webapp transfers data in a binary format using Google Protobufs, and data are compressed by Nginx before being sent over the network to the web browser. A custom JavaScript library is used on the client-side to avoid requesting duplicate data. Users login to the webapp using their database login details, and the webapp server authenticates these by attempting to connect to the database. Login sessions are persisted by storing the login details in encrypted cookies for which only the webapp server has the decryption keys. The datasets visible to the user are limited to the datasets that their database account has permission to view.

\subsection*{The Nginx server}

handles the connection between browser clients and the webapp. The server's primary function is to compress data from the web application and send it to clients. With measures in place to prevent sending duplicate data, this is an efficient way to transfer data to clients. It also serves some static content for the webapp. The Nginx server may be configured to serve the webapp using a secure protocol (HTTPS) if desired.

\subsection*{The Harvesters}

manage transferring data from the battery cyclers to the database. These are Python3 applications each running in a Docker container that is placed on the local machine connected to a battery cycler. The application's primary function is to scan appropriate predetermined locations within the local machine's file system for changes. When a change is detected, data is then parsed and it is uploaded to the Galvanalyser database. The parsing functionality of the Harvester may be customised for each type of battery test system, since they each have different file formats. At present, we have created Harvesters specific to Maccor and Ivium systems, with Biologic under development.  

The frequency at which the Harvester checks for new data can be configured to occur at whatever rate the user desires, using an appropriate task scheduler. The directories where data is sourced are mounted as volumes in the docker container. The Harvester uploads new data directly to the PostgreSQL database by accessing the database as a user.

\section*{Illustrative example}

\begin{figure}
\centering
\includegraphics[width=.9\linewidth]{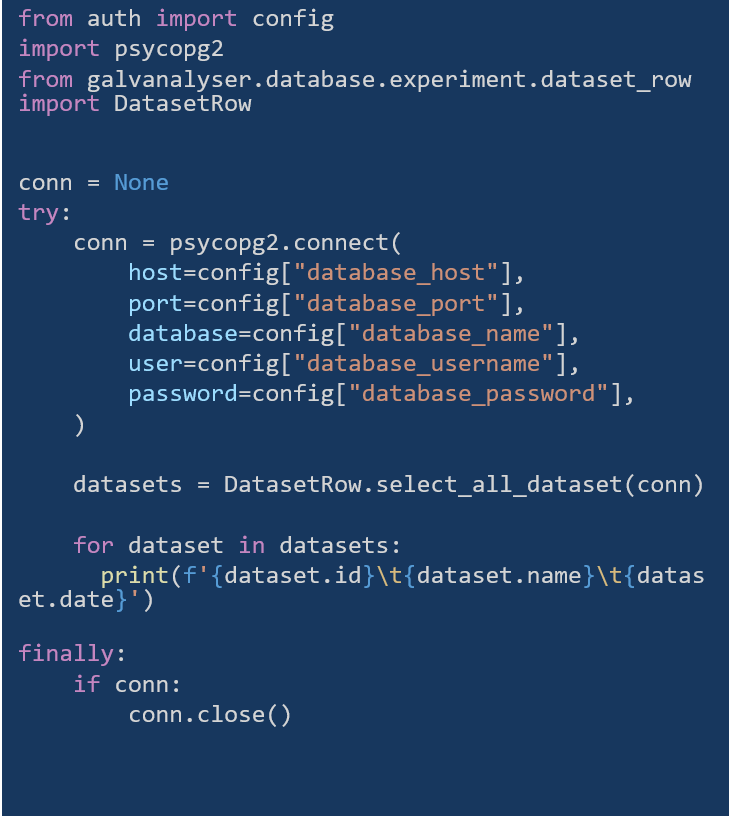}
\caption{Example Python code for accessing the Galvanalyser API}
\label{fig:GalvanalyserAPI}
\end{figure}

\begin{figure}
\centering
\includegraphics[width=.9\linewidth]{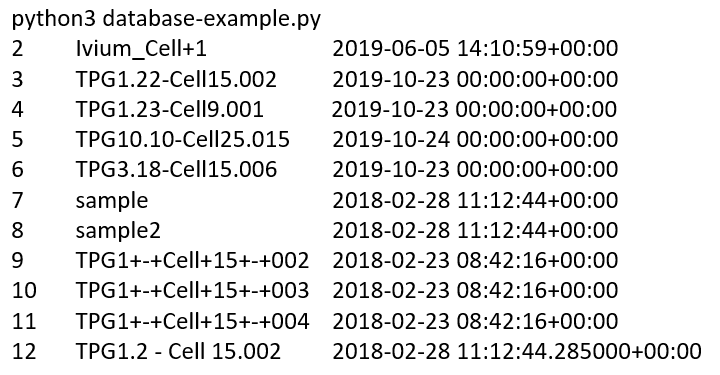}
\caption{Example of data retrieved via the API}
\label{fig:GalvanalyserAPI2}
\end{figure}

As mentioned previously, Galvanalyser may be accessed directly from various software environments, and a simple Python API is available to facilitate this. Fig.\ \ref{fig:GalvanalyserAPI} gives example Python code that connects to the Galvanalyser database through a  PostgreSQL driver, \verb|psycopg2|. The username and password required are the same as the ones used in the database. The example script shown connects to the database and requests all data that the user with the entered credentials has access to, then closes the connection after data has been transferred. The output of this request is shown in Fig.\ \ref{fig:GalvanalyserAPI2} for this particular example.

\begin{figure*}[t!]
\centering
\includegraphics[width=.85\linewidth]{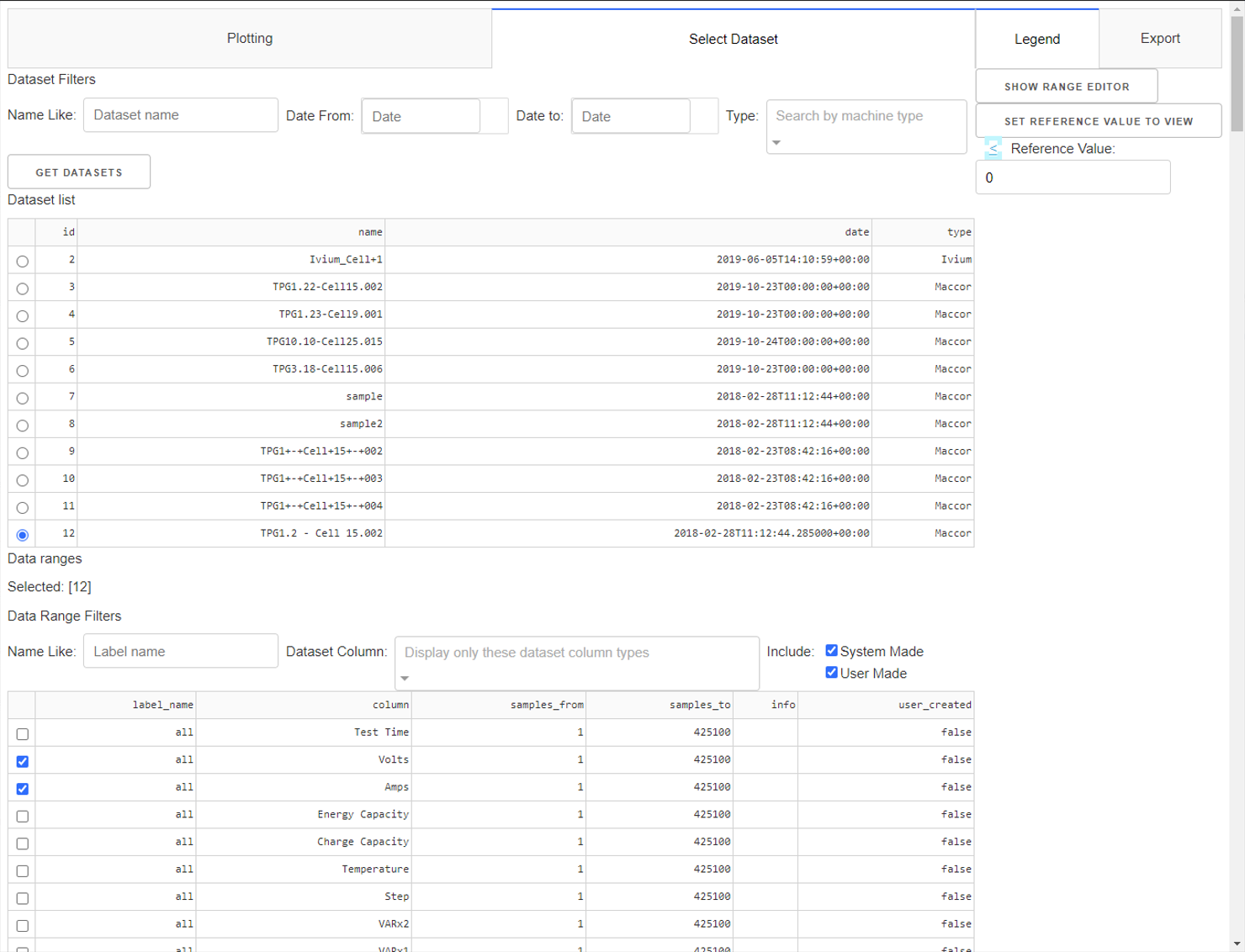}
\caption{Example of data selection using the Galvanalyser webapp}
\label{fig:GalvanalyserWebAppSelect}
\end{figure*}

\begin{figure*}
\centering
\includegraphics[width=.85\linewidth]{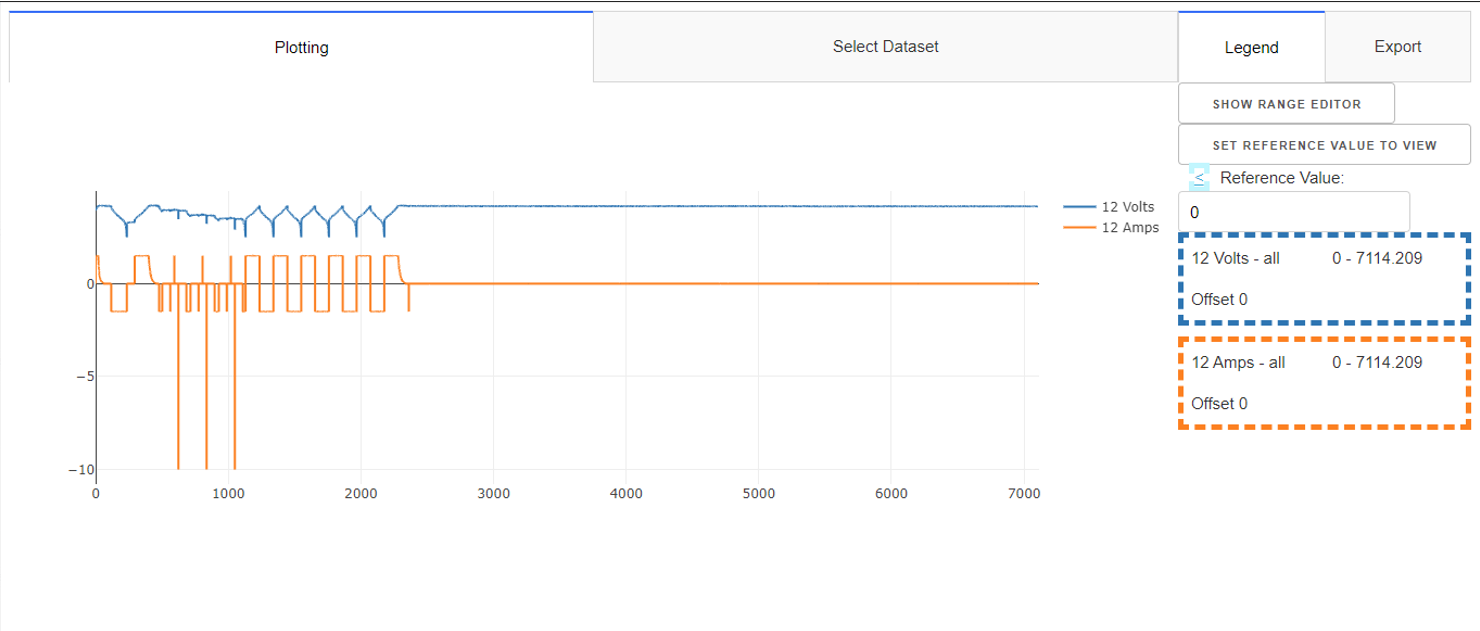}
\caption{Example of data plotting using the webapp}
\label{fig:GalvanalyserWebAppPlot}
\end{figure*}

Galvanalyser is also designed to allow a user to search for and visualise data from within a browser. Fig.\ \ref{fig:GalvanalyserWebAppSelect} shows the webapp interface for selecting data. The `Select Dataset' tab allows users to search through all the datasets available to them, by name, date range, and type. When a dataset is selected, all the data columns are then listed to allow for specific data to be exported or plotted. Fig.\ \ref{fig:GalvanalyserWebAppPlot} shows the result of plotting the selected variables, in this case time series of voltage and current, visualised using Plotly. 

\section*{Impact}

As the commercial success of batteries, particularly lithium-ion chemistries, continues to increase, the value of improving our understanding of these complex devices and accelerating efforts to learn from data goes up. Galvanalyser enables research labs to expand their battery cycler test equipment range without worrying about the growing data management problem that might arise with this. The system transfers data to a single, easily accessible location in a common searchable format. Researchers can connect remotely to the database, without having to worry about physical access to the test equipment. Our goal is to create a simple, open data management experience, that requires minimal oversight and can easily be expanded to include new users and data sources.

\section*{Conclusions}

We have presented `Galvanalyser', a database solution to the problem of managing battery test data.  Galvanalyser automatically collects data from various separate local computers connected to test equipment, parses it into a common format, then stores data in a single PostgreSQL database with row-level user access privileges. This allows for easy backups to be made, and allows users to conveniently access data through a web application or directly using an API. As a result, data may be exchanged and analysed more easily, speeding up battery design and development work. Please contact us if you want to become an early user!

\begin{acknowledgements}
We are grateful for financial support from the Pitch-In project, funded by Research England's Connecting Capability Fund (CCF). Thanks also to Dr Andy Gilchrist for facilitating this.
\end{acknowledgements}

\section*{Bibliography}
\bibliography{references_2.bib}

\begin{thebibliography}{1}
\providecommand{\url}[1]{#1}
\csname url@samestyle\endcsname
\providecommand{\newblock}{\relax}
\providecommand{\bibinfo}[2]{#2}
\providecommand{\BIBentrySTDinterwordspacing}{\spaceskip=0pt\relax}
\providecommand{\BIBentryALTinterwordstretchfactor}{4}
\providecommand{\BIBentryALTinterwordspacing}{\spaceskip=\fontdimen2\font plus
\BIBentryALTinterwordstretchfactor\fontdimen3\font minus
  \fontdimen4\font\relax}
\providecommand{\BIBforeignlanguage}[2]{{%
\expandafter\ifx\csname l@#1\endcsname\relax
\typeout{** WARNING: IEEEtran.bst: No hyphenation pattern has been}%
\typeout{** loaded for the language `#1'. Using the pattern for}%
\typeout{** the default language instead.}%
\else
\language=\csname l@#1\endcsname
\fi
#2}}
\providecommand{\BIBdecl}{\relax}
\BIBdecl

\bibitem{noauthor_voltaiq_2020}
\BIBentryALTinterwordspacing
``\BIBforeignlanguage{en-US}{Voltaiq - {Battery} {Intelligence} {Software}
  {Platform}},'' Oct. 2020. [Online]. Available: \url{https://www.voltaiq.com/}
\BIBentrySTDinterwordspacing

\bibitem{noauthor_battery_2020}
\BIBentryALTinterwordspacing
``\BIBforeignlanguage{en}{Battery {Data} {Management} {\textbar} {Astrolabe}
  {Analytics}},'' Oct. 2020. [Online]. Available:
  \url{https://www.astrolabe-analytics.com}
\BIBentrySTDinterwordspacing

\bibitem{herring_beep_2020}
\BIBentryALTinterwordspacing
P.~Herring, C.~Balaji Gopal, M.~Aykol, J.~H. Montoya, A.~Anapolsky, P.~M.
  Attia, W.~Gent, J.~S. Hummelshøj, L.~Hung, H.-K. Kwon, P.~Moore,
  D.~Schweigert, K.~A. Severson, S.~Suram, Z.~Yang, R.~D. Braatz, and B.~D.
  Storey, ``\BIBforeignlanguage{en}{{BEEP}: {A} {Python} library for {Battery}
  {Evaluation} and {Early} {Prediction}},''
  \emph{\BIBforeignlanguage{en}{SoftwareX}}, vol.~11, p. 100506, Jan. 2020.
  [Online]. Available:
  \url{http://www.sciencedirect.com/science/article/pii/S2352711020300492}
\BIBentrySTDinterwordspacing

\bibitem{samuel-buteau_samuel-buteauuniversal-battery-database_2020}
\BIBentryALTinterwordspacing
Samuel-Buteau, ``Samuel-{Buteau}/universal-battery-database,'' Sep. 2020,
  original-date: 2019-10-15T15:33:35Z. [Online]. Available:
  \url{https://github.com/Samuel-Buteau/universal-battery-database}
\BIBentrySTDinterwordspacing

\bibitem{group_postgresql_2020}
\BIBentryALTinterwordspacing
P.~G.~D. Group, ``\BIBforeignlanguage{en}{{PostgreSQL}},'' Oct. 2020. [Online].
  Available: \url{https://www.postgresql.org/}
\BIBentrySTDinterwordspacing

\bibitem{noauthor_empowering_2020}
\BIBentryALTinterwordspacing
``\BIBforeignlanguage{en}{Empowering {App} {Development} for {Developers}
  {\textbar} {Docker}},'' Oct. 2020. [Online]. Available:
  \url{https://www.docker.com/}
\BIBentrySTDinterwordspacing

\end{thebibliography}

\appendix

\section{Database structure}
\label{appendix}

\begin{figure*}[ht!]
\centering
\includegraphics[width=.8\linewidth]{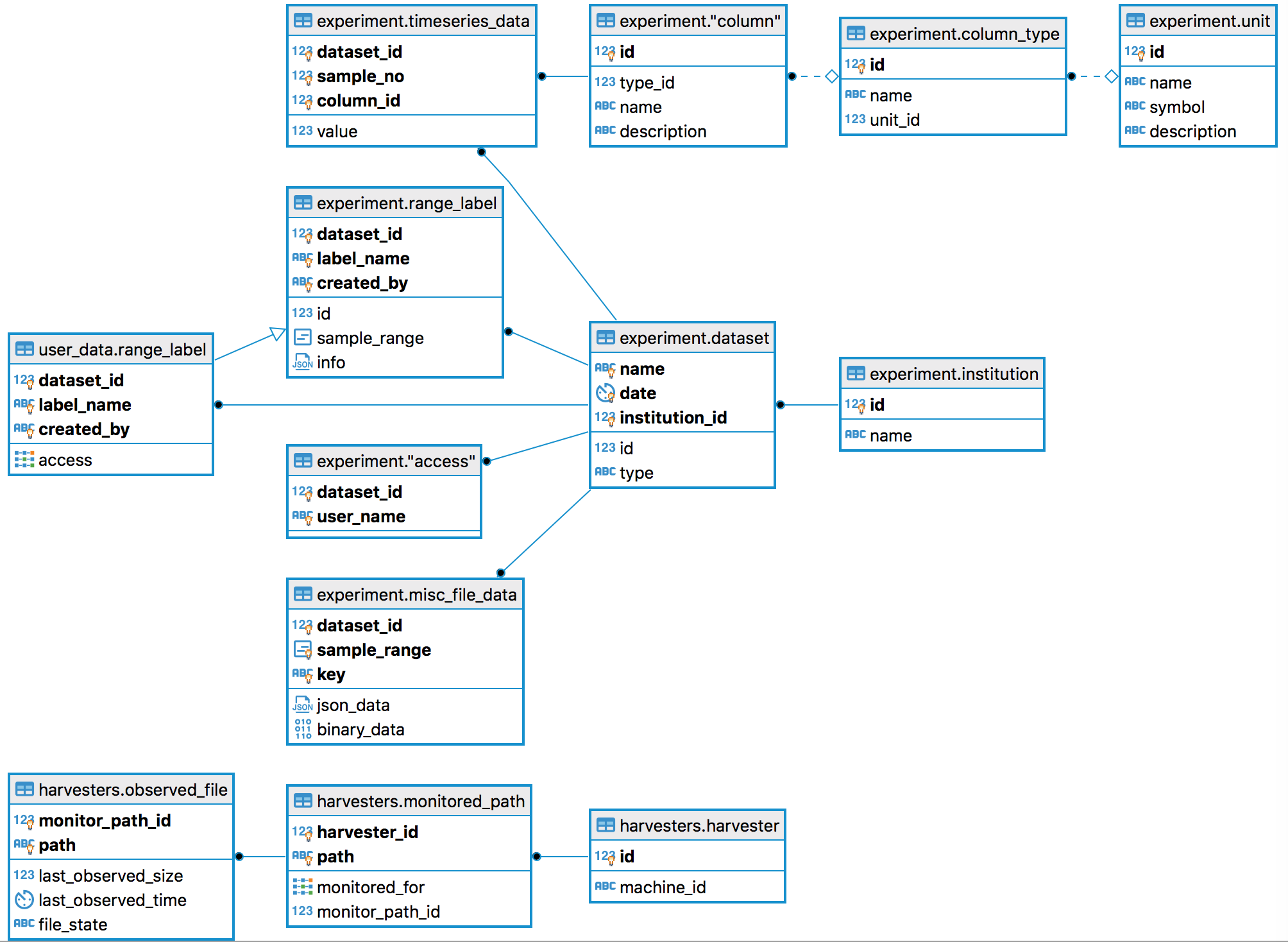}
\caption{Galvanalyser PostgreSQL database entity relationship diagram}
\label{fig:GalvanalyserERD}
\end{figure*}

Fig.\ \ref{fig:GalvanalyserERD} shows the entity relationship diagram for the Galvanalyser system. 

A dataset is defined by \verb|experiment.dataset| which then has a one-to-many relation to \verb|experiment.timeseries_data|. This entity uses \verb|sample_no| to identify samples, rather than sample time, because some cyclers don’t output time with sufficient resolution for timestamps to be unique between samples, whereas an integer that increments with each row of samples is easy to generate if required and is guaranteed to be unique.

The \verb|experiment.timeseries_data| table is structured to allow it to store any number of columns of whatever floating point data is required; this is achieved through the entity relation shown to the right of this table. This method provides flexibility but avoids trying to make a table with every column that might ever be needed, which in practice would end up having lots of NULL values, since each log file would only ever contain a subset of the data.

The \verb|experiment.column| table also has an entity relation to \verb|experiment.column_type|. This allows association of column types, and units attached to those types, without duplicating data. For example, this allows columns \verb|channel 1 voltage|, \verb|channel 2 voltage| and \verb|channel 3 voltage| to be `voltage' type columns, and `voltage' columns to have units of volts.

The entry \verb|experiment.dataset| also has a many-to-one relation with \verb|experiment.institution|. This allows data from different institutions to be uploaded to a common database if required. The entry \verb|experiment.misc_file_data| allows a \verb|sample_range| to be defined and used store any other data from the log files that does not fit in any of the other tables.

The \verb|harvester.*| tables exist to let the Harvesters know where to look for files, who those files belong to, what the state of the files are. Finally, the \verb|harvesters.monitored_path| defines a file-system path to a specific folder, with a one-to-many relation to \verb|harvesters.observed_path|, which contains specific information to each file the harvester is monitoring.

\end{document}